\documentstyle[prl,aps,12pt]{revtex}
\begin{document}
\title{Classical Diffusion and Quantum Level Velocities: Systematic
Deviations from Random Matrix Theory}
\author{Arul Lakshminarayan$^{\ast}$, Nicholas R. Cerruti and Steven Tomsovic}
\address{Dept. of Physics, Washington State University,\\
Pullman, WA, 99164 USA 
\footnote {Permanent Address: Physical Research Laboratory,
Navrangpura, Ahmedabad 380 009, India.}} 
\maketitle
\newcommand{\newc}{\newcommand}
\newc{\beq}{\begin{equation}}
\newc{\eeq}{\end{equation}}
\newc{\kt}{\rangle}
\newc{\br}{\langle}
\newc{\longra}{\longrightarrow}
\newc{\eps}{\epsilon}
\begin{abstract}
We study the response of the quasi-energy levels  
in the context of quantized chaotic systems through the 
level velocity variance  and relate them 
to classical diffusion coefficients using detailed semiclassical
analysis. The systematic deviations from random matrix theory, assuming
independence of eigenvectors from eigenvalues, is shown to be connected to
classical higher order time correlations of the chaotic system. 
We study the standard map as a specific example, and 
thus the well known oscillatory behavior of the diffusion
coefficient with respect to the parameter 
is reflected exactly in the oscillations of the variance of the level 
velocities. We study the case of mixed phase-space dynamics as well 
and note a transition in the scaling properties of the variance 
that occurs along with the classical transition to chaos. 
\end{abstract}

\newpage

\section{Introduction}

The quantum spectrum is well known to reflect in several ways 
classical integrability or its lack thereof \cite{Gutzbook,LesHou91,Almbook}. 
For a completely chaotic, quantized  system
the energy eigenvalues have characteristic,
in fact, universal fluctuation properties that coincide with  random 
matrix theory (RMT) universality classes and the eigenfunction 
components are also distributed as Gaussian random variables. However, there
are important deviations from this dull uniformity imposed by the underlying  
(asymptotic) deterministic chaos. Classical periodic orbits, a dense set of 
measure zero unstable orbits, introduce characteristic deviations that
are well documented, including the phenomenon of eigenfunction scarring 
\cite{Heller}.
The movement of energy levels with the variation of an external parameter,
level dynamics, has also been studied by several authors with different
motivations \cite{GRN89,TakHas,ZakDel,SimAlt,Sano,Kunst}. 
It is known that the motion of the energy levels 
as a function of the parameter,
now a psuedo-time variable, is  completely integrable  whether the system is 
itself chaotic or not \cite{NakLak}. Nevertheless 
there are characteristic features that are introduced by 
chaos, for instance  avoided crossings that may be characterized by the
second derivative of the energy levels, {\it i.e.} the curvatures. 

	Here we study  level ``velocities'', and relate them
directly to classical diffusion coefficients; although we are using the 
term velocities, we are not discussing adiabatically changing a system, 
just the slopes of the level curves as a function of a controllable 
parameter.  It has been known for some time
that these are Gaussian distributed with a variance that
has been related to a classical ``generalized conductance'', especially in
the context of weakly disordered metallic grains.   
Methods employed were mostly field theoretic and RMT based, while
numerical simulations of chaotic billiards led   
to  the conjecture  that the behavior of disordered systems could
be extended to chaotic ones as well \cite{SimAlt}. 

The variance has a significance beyond setting the scale of the 
Gaussian distribution of velocities. It enters as a normalization required to 
uncover possible universalities in parametric level correlations.
It encodes the system specific
characteristics of level motions as a function of an external parameter. 
Level correlations and velocities  are experimentally accessible, 
for example in microwave cavities \cite{Sridhar} or 
quantum dots, 
Although, universal parametric correlations are not well established
experimentally, a recent experiment exploiting the 
similarity of elastomechanical 
wave equations of flexural modes of plates to the Schrodinger equation, 
seems to lend support to it \cite{SchadKud}.
    
In the case when the changing parameters are Aharanov-Bohm 
flux lines that do not lead to any classical dynamical changes, but do 
lead to important spectral modifications, correlation between level velocities
were semiclassically considered in \cite{BerrKeat}. 
For a treatment of Hamiltonian flows see \cite{Nick}.
Recent closely related
work, in the context of Hamiltonian flows, is also found in \cite{LebSieb},
where detailed results about the variance of level velocities are presented  
for billiards. 

 We make precise the connection between classical diffusion and the variance of
the level velocities in the simpler context of quantized maps or more 
generally time periodic systems where detailed semiclassical (and classical)
analysis is possible. We evaluate the variance for the standard map as a
function of the external kicking strength and show system specific correlations
in the form of Bessel function oscillations. Since two-dimensional 
area preserving, or more 
generally symplectic maps, are Poincare sections of
Hamiltonian flows, our analysis also reflects upon these systems and is consistent
with  results derived therein. On the other hand, due to the vastly simpler 
numerical and analytical work involved with maps, they lend themselves to more 
detailed and extensive work.

We relate our analysis to a semiclassical evaluation of expectation
values of generic operators in the eigenbasis, as well as touch upon 
two parameter variations and their correlations. The case when the dynamics 
leads to a  mixed phase space is generic and we find a 
Weyl type expansion in $\hbar$ for the variance. The principal 
contribution in this regime is well predicted by a simple {\it classical}
correlation, which vanishes as the system undergoes a transition to chaos.
The different scaling behaviors in effective $\hbar$ 
for mixed and chaotic systems can be 
experimentally observed. We will consider the standard map as an example. 
Others before us have used such systems to study level dynamics 
\cite{Sano,SahHG}.

\subsection{The Standard Map and Random Matrix Theory}
Here we define the model studied below and derive the RMT predictions 
for these. 
Let the classical Hamiltonian have the form
\beq
\label{kickhamilt}
	H=p^2/2 \, -\, \lambda V(q) \, \sum_{n=-\infty}^{\infty}
\delta(t-n)
\eeq
so that the Floquet operator connecting states just before kicks is
given by 
\beq
\label{floquetop}
	U\, =\, \exp(-i p^2/2\hbar) \, \exp(i \lambda V(q)/\hbar).
\eeq
The time between kicks is taken to be unity, as there are two independent
parameters already present, namely $\lambda$ and $N$.
Such systems, known as quantum maps, 
were first studied in \cite{CCIF,BBTV} and led to the 
uncovering of dynamical localization, akin to Anderson localization in
disordered conductors \cite{Prange}.  
 We will typically consider the
above to be the way the parameter of interest ($\lambda$) enters the problem.

While this is a map on the plane (for one-degree-of-freedom systems),
we consider their restriction to the torus $[0,1)^2$. 
This is essential as we have in mind 
bounded Hamiltonian systems and not open scattering ones. Periodic boundary 
conditions are imposed in both $p$ and $q$ directions. 
We will assume that $V(q)$ is 
a smooth function on $[0, 1)$ with unit periodicity. Denote its average as
\beq
\label{averageV}
	\overline{V}\, =\,\int_{0}^{1} V(q) \, dq.
\eeq
Let the quantum map be the $N$ dimensional unitary matrix operator denoted
by $U$. Maps, such as the standard map,  restricted to a  torus are 
quantized using standard canonical quantization \cite{Izrailev}. 
Periodic boundary 
conditions in both canonical variables imposes a finite number of states which
is the inverse effective Planck constant $(h=1/N)$. 
Thus the classical limit
is approached in the large $N$ limit. Various quantum maps on the torus 
have been studied and form an important part of the literature on quantum 
chaos due to their inherent simplicity \cite{AndrePRL,BVbak,ShiChang,HB}.
The discrete spectra ($N$ levels) 
obtained are then analyzed for various properties, in particular here the
eigenangle velocities are obtained. 

The classical standard map is given by
the recursion 
\begin{eqnarray}
\label{classmap}
q_{i+1}&=& (q_{i}+p_{i+1}) \, \mbox{mod}\,  (1)\\ \nonumber
p_{i+1}&=&(p_{i}-(k/2 \pi) \sin(2 \pi q_{i})) \, \mbox{mod}\,  (1),\nonumber  
\end{eqnarray}
where $i$ is the 
discrete time. This is the solution to the Hamiltonian equations of motion 
for the potential $V(q)\, =\, \cos(2 \pi q)$ and the Hamiltonian in Eq. (\ref{kickhamilt}). The dynamical variables are monitored just before the kicks,
and $\lambda=k/(2 \pi)^2$.
The standard map is of central importance as many other maps are locally 
described by this and the  potential may be considered to be
the first term in the Fourier expansion of more general periodic potentials.
The parameter $k$ is of principal interest and it controls
the degree of chaos in the map, a complete transition to ergodicity is 
attained above values of $k \approx 5$, while the last rotational KAM torus
breaks around $k \approx .971$ \cite{Reichl}. 

The quantum map in the discrete position basis is given by \cite{LakPRM}
\beq
\label{quantummap}
      \br n |U| n^{\prime} \kt \, =\, \frac{1}{\sqrt{i N}} 
\exp\left(i \pi (n-n^{\prime})^2/N\right) 
\exp \left(i \frac{k N}{2 \pi} \cos(2 \pi (n+a)/N) \right).
\eeq
The  parameter to be varied will be the ``kicking strength'' $k$, while the phase
$a$ will be used to avoid exact quantum symmetries, and $n, n^{\prime}\, =\, 
0, \ldots, N-1$. 
The eigenvalue problem of the unitary matrix 
is written as $ U |\psi_j 
\kt  \, =\, \exp(-i \phi_j) |\psi_j \kt $. The eigenangles $\phi_j$
are real and their variation with the parameter $k$ (level ``velocities'')
are  given simply by
the matrix elements
\beq
\label{velomatele}
\frac{d  \phi_j}{d  k} \, =\,  \frac{N}{2 \pi}  \br \psi_j |V|\psi_j \kt \, =\, 
 \frac{N}{2 \pi} \br \psi_j |\cos(2 \pi q)|\psi_j \kt.
\eeq
The $2 \pi$ factor is the result of choosing $k$ as the relevant parameter
and not $k/2 \pi$ and we retain this as this corresponds to the more 
conventional usage where the last KAM torus breaks when the parameter
value is just under unity.

It is then clear that studying level velocities is equivalent to studying
expectation values of operators in the eigenbasis. Thus if we require 
$\br \psi_i |A| \psi_i \kt $ we would look at the modified unitary operator
(assuming $A$ is Hermitian)
\beq
\label{operator}
	U\, =\, U_0 \exp(-i \lambda A/\hbar)
\eeq
where $U_0$ is the quantum system under study. Then the expectation values
are simply the corresponding level velocities evaluated at 
$\lambda=0$, multiplied by $\hbar$.
If one may identify the classical canonical transformation generated by 
$A$, we could  study a modified classical map, as well. However since 
$\lambda=0$, it is the properties of the original classical map
that will be relevant. The work \cite{EFMW92} already discussed the general
problem of semiclassical evaluation of matrix elements and our following
work may be viewed in this context as well. 
 
From the Gaussian distribution of eigenfunctions for a quantum chaotic system
we expect the level velocities be similarly distributed. We will
concentrate on the variance of these velocities, namely the sum:
\beq
\label{variance}
\sigma^2(k, N) \, =\, \frac{1}{N} \sum_{j=1}^{N}\left
(\frac{d  \phi_j}{d  k}\right)^2 -
\left(\frac{1}{N} \sum_{j=1}^{N}\left(\frac{d  \phi_j}{d  k}\right) \right)^2,
\eeq     
We will assume, as is the case with the standard map example, that the
average vanishes, {\it i.e.} $\overline{V}\, =\, 0$.
Later we will generalize to the case of a non-vanishing
average, or expectation values of operators with non-zero traces.
Figure 1 shows a scaled $\sigma^2$ as a 
function of the parameter $k$. At about $k \approx 5$ the variance settles
down to a near constant, this value coincides with the disappearance
of major islands of stability in the classical phase space. What interests
us primarily here however is the clear oscillations that persist as a function
of $k$ right into regions of large chaos as shown in the inset.

First we study the value around which the oscillations occur, as this is
provided by assuming RMT models.
Using Eq.~(\ref{velomatele}) 
we get 
\begin{eqnarray}
\label{rmt1}
\sigma_{\mbox{RMT}}^2 & =& \frac{N}{4 \pi^2}  \sum_{m=0}^{N-1} \left( \sum_{n=0}
^{N-1} \left| \br \psi_m |n \kt \right|^2 \, V((n+a)/N) \right)^2 
\nonumber\\
&=& \frac{N}{4 \pi^2}  \sum_{m=0}^{N-1} \left( \sum_{n=0}^{N-1} 
\left| \br \psi_m |n \kt \right|^4 \, (V((n+a)/N))^2 \right. +\\
&& \left. \sum_{n \neq n^{\prime}}
\left| \br \psi_m |n \kt \right|^2 \left| \br \psi_m |n^{\prime} \kt \right|^2 
\, V( (n+a)/N)  V( (n^{\prime}+a)/N) \right) \nonumber.
\end{eqnarray}

The eigenfunctions have been expanded in a basis that diagonalizes the 
perturbation $V$, which we have taken to be the position basis.
Since we assume a zero centered or trace-less
pertubation 
\[
\sum_{n=0}^{N-1} V((n+a)/N) \, =\, 0.
\]
We use the square of this relation in Eq.~(\ref{rmt1}) while
replacing  eigenfunction components by their ensemble averages (denoted  by
angular brackets) to derive that
\beq
\label{rmt2}
\sigma_{\mbox{RMT}}^2 \, =\, \frac{N^3}{4 \pi^2}  \overline{V^2}
\left( \left< |\br \psi_m|n \kt|^4  \right> 
\, -\,  \left< |\br \psi_m|n \kt|^2  |\br \psi_m|n^{\prime} \kt|^2 \right> 
\right).
\eeq
A crucial step in writing down the above is to assume the independence
of the eigenfunction components from any specific position eigenvalues. 
While this is a reasonable statistical assumption we will see below that
it misses important correlations that are incorporated naturally in 
semiclassical treatments. This is the origin of the 
non-universality of level dynamics, as this implies system dependent
correlation effects. The same perturbations ($V$) applied to different chaotic
systems will result in different statistical responses, 
unlike the predictions of RMT.
 
We use standard results from RMT relevant to the 
Gaussian Orthogonal Ensemble (GOE),
which is applicable here as well
as the relevant Circular ensembles, \cite{PandeyRMP}. In particular 
\[ \left< |\br \psi_m|n \kt|^4  \right> \, =\, \frac{3}{N(N+2)} \sim \frac{3}
{N^2}, 
\; \; \left< |\br \psi_m|n \kt|^2  |\br \psi_m|n^{\prime} \kt|^2 \right> 
\, =\, \frac{1}{N(N+2)} \sim \frac{1}{N^2}.\]
 We finally get
\beq
\label{rmt3}
\sigma_{\mbox{RMT}}^2 \, =\, \frac{N}{2 \pi^2}  \overline{V^2}.  
\eeq
As a special case for the standard map $\overline{V^2}=1/2$ and we get
$\sigma_{\mbox{RMT}}^2 \, =\, N/4 \pi^2$. 
This last result explains the value  about which the oscillations occur 
in Fig. 1. This implies  that the response of the system as measured
by the movement of the energy levels is essentially the intensity of the 
perturbation. For chaotic systems then the response is 
 independent of the system's  
detailed dynamical properties. 
We must also point out that when time reversal symmetry is 
broken the response is half as large.
We now turn to the systematic oscillations that are not 
readily predicted by random matrix theory and are manifestly
system dependent.

\section{Semiclassical Theory}

\subsection{The chaotic phase-space}

We first develop in some generality expressions for the variance of the 
level velocities in
which semiclassical methods can be easily applied. We write a gaussian
smoothed density of states \cite{AlmSara} as 
\beq
\label{smoothrho}
\rho_M(\phi) \,=\, \sum_{n=-\infty}^{\infty} \, F_M(n)\; \exp(i n \phi) \; \mbox{Tr}\, U^n,
\eeq
where $F_M(n)\,=\, \exp(-n^2/2M^2)/(2 \pi)$ is introduced to avoid divergences. The exact spiked density of states is obtained in the limit $M \rightarrow \infty$
although almost all levels will be resolved at $M=N$, as the mean level 
spacing is $2 \pi/N$. The smoothed step function, $N_M(\phi)$ is the integral of the level density with respect to $\phi$. We derive then that 
\beq
\label{numberden}
\int_0^{2 \pi} \left(\frac{dN_M(\phi)}{dk} \right)^2 \, d \phi \, =\, 
\frac{M N \sigma^2(k, N)}{2 \sqrt{\pi}} \, =\, 
2 \pi \, \sum_{n=-\infty}^{\infty} \frac{F_M^{2}(n)}{n^2} 
\left| \frac{d}{dk} \mbox{Tr}(U^n) \right|^2.
\eeq
The term $n=0$ does not belong in the sum, and it
is understood that the first equality is an approximation that becomes exact as
$M \rightarrow \infty$. From this expression it follows that it is the
long time traces of the propagator, and therefore semiclassically, long 
periodic orbits that are important.

Another very similar route is through the identity
\beq
\mbox{Tr}\left( U^n V \right) \, =\, \sum_{j=1}^{N-1} \br \psi_j |V|\psi_j \kt
\exp(-i \phi_j n)
\eeq
thus implying that 
\beq
\label{sigmatr}
\sigma^2(k,N)\, =\, \frac{N}{4 \pi^2} \left< | \mbox{Tr}\left( U^n V \right)
 ^2 \right>_n,
\eeq
where the angular brackets indicate averaging over time $n$ in the 
neighborhood of large $n$. We assume as is
relevant for chaotic systems that there are no degenerecies. To make connections with the standard map above we would take $V\, =\, \cos(2 \pi q)$.

Now we make use of the semiclassical approximation of the trace of the 
propagator as a sum over periodic orbits \cite{Gutzbook,Tab83}
which is 
\beq
\label{trace1}
\mbox{Tr}\left( U^n \right) \sim n \sum_{p} A^{(n)}_p 
\exp\left( 2 \pi i N S^{(n)}_p \, -\, i \pi \nu_p/2 \right),
\eeq
where $A^{(n)}_p\, =\, 1/(2 \sinh(\lambda_p n/2 ))$, and 
the sum is over periodic orbits of period $n$ which labelled by $p$ and
have a Lyapunov exponent $\lambda_p$. The actions of these orbits
are denoted by $S^{(n)}_p$ and are calculated from the
generating function of the classical map. The phases $\nu_p$ are 
Maslov like indices and will not concern us here.

There is also a generalization of the above, which is particularly
easy to derive when the perturbation is diagonal in the position 
(or momentum) basis. 
\beq
\label{trace2}
\mbox{Tr}\left( U^n V\right) \sim  \sum_{p} A^{(n)}_p 
\exp\left( 2 \pi i N S^{(n)}_p \, -\, i \pi \nu_p/2 \right) \sum_{j=1}^{n}
V(x_j^p).
\eeq
Here $V(x_j^p)$ is the value of a phase space representation of the operator
$V$ that is evaluated along the periodic orbit labelled $p$ and at the point
labelled $j$. 
An appropriate generalization in the energy
domain for continuous time systems is found in \cite{EFMW92}.
The sum around the periodic orbit of the function $V$ is essentially the
derivative of the action with the parameter, and we may use either the first
trace formula in conjunction with Eq.~(\ref{numberden}) or the second
with Eq.~(\ref{sigmatr}).

We take the second route as we connect with the first subsequently.
Taking the modulus of the second trace formula gives
\begin{eqnarray}
\label{modtraceV}
|\mbox{Tr}\, (U^n V)|^2 & \sim & 
 \sum_{p} A_p ^{(n)2} (\sum_{j=1}^{n} V(q_{j}^p))^2 \, +\, \nonumber \\ 
& &\sum_{p \ne p^\prime} A_p^{(n)} A_{p^\prime}^{(n)} \,
(\sum_{j=1}^{n} V(q_{j}^p) )(\sum_{j=1}^{n} V(q_{j}^{p{^\prime}})) 
\exp(2 \pi i N (S_p - S_{p^{\prime}})).   
\end{eqnarray}
As is usual, we have separated the diagonal contribution from the
``off-diagonal'', which corresponds to distinct pairs of orbits, with
distinct actions.  We have also
assumed for simplicity, as is the case with the specific parameter variation
chosen above in the standard map, that $V(x)$ is only position dependent; this
does not alter the results below. 
We have also included the phases into the actions.

Since we expect that long 
periodic orbits are important, the diagonal approximation, which relies on
random phases may be violated due to subtle correlations among their actions.
The time at which we may expect 
action differences of the order of $\hbar$ is the so called log-time, or
Ehrenfest time. We argue that action differences are of the order of
the orbit separation, and since areas of the order of $\hbar$ 
(for two-dimensional maps) would be populated with multiple periodic orbits beyond the log-time,
their action differences would also be comparable with $\hbar$. 
However long periodic orbit actions are randomly distributed and will acquire
correlations only around the Heisenberg time. At this time the off-diagonal 
terms will dominate the sum, as happens if we simply consider
$<|\mbox{Tr}\, (U^n)|^2>_n$, which is asymptotically  $N$,  
while the diagonal term is linearly increasing in time.  

However the off-diagonal term vanishes due to the sums of $V(q)$ over 
very long periodic orbits. 
We may write for two distinct orbits $p$ and $p^{\prime}$
after  assuming uniform measure
and replacing time averages over the periodic orbit, by the phase-space 
average that  
\beq
\label{vjvjp}
(\sum_{j=1}^{n} V(q_{j}^p) )(\sum_{j=1}^{n} V(q_{j}^{p{^\prime}}))
\sim n \, \overline{V}^2
\eeq
From our initial assumption that $\overline{V}\, =\, 0$, the  
off-diagonal term vanishes.
The diagonal term is non-vanishing as we once again treat periodic
orbits as ordinary long chaotic trajectories and derive that 
\beq
\label{correlD}
\frac{1}{n}(\sum_{j=1}^{n} V(q_{j}^p))^2 \equiv  \, D(k)= \, C(0)+2\, \sum_{l=1}^{n}C(l)
\eeq
where the time correlations are replaced by classical phase space averages
due to ergodicity.
\beq
\label{correl}
C(l)\, = \, \left< V(q_0) V(q_{l}) \right>\, =\, \frac{1}{{\mathcal A}} 
\int_{{\mathcal A}} dq_0 \, dp_0 V(q_0) V(f^l(q_0,p_0)),
\eeq
where $f^l(q_0,p_0)=q_l$, $f^l$ is  the integrated dynamics in time $l$, 
and ${\mathcal A}$ denotes both the phase space and  its
area (in the cases considered this is unity).
We assume that these exist, and are decreasing with $l$, 
typically exponentially for chaotic systems and  that a few terms may be 
sufficient. This is not established in generality and complications 
may arise due to marginally stable orbits leading to non-exponential 
behaviors. For the standard map, coefficients upto $C(2)$ are dominant and 
sufficient to see the essential behaviour.
We have dropped the index $p$ as now we will treat such long 
periodic orbits as generic non-periodic orbits. Indeed by using the ergodic
theorem we have already abandoned any particularities that may arise 
due to the orbit being periodic. 
Later, we remark on a case
when we may not neglect off-diagonal terms. 
 
The alternative route is to take the derivative of the first trace formula
and use Eq.~(\ref{numberden}). Again we neglect the off-diagonal terms for
reasons given above.
We will assume that the derivatives of the actions with the parameter 
(``action velocities'')
, for a given period or period interval,  
are such that their average is zero while their variance is 
proportional to the time period. This assumption is equivalent to the 
vanishing of the phase space average of $V(q)$ and  the 
presence of  ergodicity.  
This was noted for general Hamiltonian systems in 
\cite{BGAS},
and we will see below in the context of maps how this simply arises.   
We replace then for each time $n$ the individual action velocities (squared)
by the variance, 
\beq
\label{actiondiff}
\left< \left(\frac{dS_{p}^{(n)}}{dk}\right)^2 \right>_{p} \, =\, D(k)\, n,
\eeq
where the angular brackets indicate the average over periodic orbits of
period $n$.

In either approach, the uniformity principle \cite{HannAlm} is applied in 
the form that there are
$e^{hn}/n$ orbits each with a Lyapunov exponent approximately $\lambda$
per unit time. Then $|A_{p}^(n)|^2 \approx e^{-\lambda n}$, and assuming
near equality of the topological entropy $h$ and $\lambda$, we 
derive from Eq.~(\ref{modtraceV}) that
\beq
\label{traceVdiff1}
|\mbox{Tr}\, (U^nV)|^2 \, \sim\,  g D(k).
\eeq
Similarly from the other approach 
\beq
\label{traceVdiff2}
\frac{1}{n^2} \left|\frac{d}{dk}\mbox{Tr} U^n \right|^2 
\sim  g \,\frac{N^2}{4 \pi^2}  D(k).
\eeq

The tilde sign in the above equations implies that the L.H.S can be 
expected to be the R.H.S in an average sense. The spread in time $n$ will
also reflect the spread in the average action velocity diffusion  
coefficent $D(k)$ with period. Results not shown here indicate that in 
the chaotic regime this is an exponential distribution. 
The factor $g$ inserted above is due the fact that symmetries can impose
distinct orbits to have identical actions. This factor must be determined 
from classical and quantal symmetries, and includes phase-space symmetries.    
We finally get then from either approach the response in terms of the
variance of the level velocities:
\beq
\label{finalform}
\sigma^2(k,N)\, =\,  g \, \frac{N}{4 \pi^2} \, D(k). 
\eeq  

Thus the variance of the level velocities is proportional to a classical
diffusion coefficient that determines the diffusion of action velocities 
of periodic orbits. More explicit expressions for this coefficient are
now derived. 
Area preserving maps, such as
the standard map, have a generating function $L(q_{i+1}, q_{i}; k)$ from which
the map may be derived as $\partial L/\partial q_i \, =\, -p_i$ and 
$\partial L/\partial q_{i+1} \, =\, p_{i+1}$ (I. C. Percival in 
\cite{LesHou91}). The total action of a 
periodic orbit is equal to $S_p^{(n)}\, =\, \sum_{i}L(q_{i+1}, q_{i}; k)$,
where the sum is over the $n$ periodic points of the orbit $p$. Thus we
derive, after assuming that the orbit is not at a point of 
bifurcation,  that for a periodic orbit:
\beq
\label{actionderiv}
\frac{dS_{p}^{(n)}}{dk}\, =\, 
\frac{\partial S_{p}^{(n)}}{\partial k}.
\eeq
We have not used the partial derivative sign in defining the level velocities
although we assume that only one parameter is varied. This is due to the 
subsequent fact that when the classical action derivative is written, 
it is a total derivative,  
in as much as the periodic orbit itself changes with the parameter. These
two, however, are shown to be equal in the case of periodic orbits. 

The variance is given by 
\beq
\label{standmapvar}
\left< \left(\frac{dS_{p}^{(n)}}{dk} \right)^2 \right>_{p} \, =\, 
\left< \sum_{i,j=1}^{n} \cos(2 \pi q_i) \cos(2 \pi q_j)\right>_{p} \, \sim\, 
\frac{n}{2} \, (1+2 J_{2}(k)) 
\eeq
In the  equality the sum is over different times along a given
periodic orbit and then averaged over all periodic orbits of period
$n$, while the approximation arises from a replacement of the average by the
usual ensemble average and retaining up to the second order time correlation.
More precisely $C(0)=1/2$, $C(1)=0$, $C(2)=J_{2}(k)/2$, where $J_{2}(k)$ 
is a Bessel function.
These are derived from:

\begin{eqnarray}
\label{classbessel}
C(0) & =& \frac{1}{n} \sum_{i=1}^{n} \cos^{2}(2 \pi q_i)  \sim  
\int_{0}^{1}  \int_{0}^{1} 
\cos^{2}(2 \pi q) \, dq \, dp  \, =\,  1/2, \nonumber\\
C(1)&=& \frac{1}{n} \sum_{i=1}^{n} \cos(2 \pi q_i) \cos(2 \pi q_{i+1}) 
\nonumber\\
&  \sim & \int_{0}^{1} \int_{0}^{1} \cos(2 \pi q) \cos(2 \pi (q +p -
(k/2 \pi) \sin(2 \pi q))) \, dq \, dp \, = \,  0, \\
C(2)&=& \frac{1}{n} \sum_{i=1}^{n} \cos(2 \pi q_{i-1}) \cos(2 \pi q_{i+1}) 
\nonumber \\
&   \sim & \int_{0}^{1} \int_{0}^{1} \cos(2 \pi (q-p)) \cos(2 \pi (q +p -
(k/2 \pi) \sin(2 \pi q))) \, dq \, dp \,  = \, J_{2}(k)/2. \nonumber
\end{eqnarray}

The symmetry factor is $g=2$, 
for the standard map, if we assume time-reversal invariance alone.
This is the case for the data presented in Fig. 1 as we have intentionally
broken the phase space symmetry in the quantum system by assuming $a=.35$,
(generic value) rather than $a=.5$, which will lead to twice the variance.
Periodic orbits are either self-symmetric or
more generically have symmetric partners with identical actions.

The Bessel functions are  characteristic of the diffusion coefficient
in the standard map, the above  simple derivation was proposed 
in the context of deterministic diffusion in  \cite{ChirikovRPM}.
The linear time dependence is then a consequence of ergodicity, whereby
time averages are replaced by phase space averages, and the coefficient is
also easy to find and is the scaling of the parameter introduced to
uncover possible universalities in level dynamics.  
These then are the oscillations observed in the figure. 
The bold line in the inset is $2 D (k) \, =\, (1+2 J_{2}(k))$. 
The significant
deviation around the first minimum in the inset ($k \approx 6.5$) 
from theory could be due to 
the presence of small stable islands, which are the accelerator modes and
are known to lead to anomalous transport in the standard map (B. V. Chrikov 
in \cite{LesHou91}). 

\subsection{The mixed phase-space}

The regime where there is a mixed phase-space consisting of large stable
regions is generic, and in this case the analysis above fails: the assumptions
about the trace formula and the uniformity principle operate only under
conditions of complete hyperbolicity. 
While in the completely chaotic regime the variance scales as $N$, in the
mixed phase space regime it (principally) scales as $N^2$. 
This relates to the large hump 
in Fig. 1, to which we now turn our attention. One way of relating the
variance to classical quantities is to recognize  that
\beq
\label{tracecorrel}
\sigma^2(k,N)\, =\, \frac{ N}{4 \pi^2} 
\left< \mbox{Tr}(U^{-n} V U^n V) \right>_n,
\eeq
where the average is once more the time average. Thus the variance of 
the quantum level velocities is directly the time average of operator
auto-correlations. We consider the case when the average level velocity 
is zero, as the generalization is evident. 
Replacing operators by the corresponding classical
observables, we expect to get the variance in the mixed phase regime.
This is particularly successful as we are dealing with averages over the
entire quantum spectrum. Thus we replace the trace operation divided by $N$
with the classical phase-space average to get
\beq
\label{classicalvel}
\sigma^2(k,N)_{cl}\, = \, \frac{N^2}{4 \pi^2} \int_{0}^{1} dp_0 \, 
\int_{0}^{1} dq_0
 \, V(q_0)\, \left< V(q^{(n)}(q_0,p_0))\right>_n,
\eeq
where $q^{(n)}(q_0,p_0)$ is the position after $n$ iterations starting from
the initial condition $(q_0,p_0)$.
For the case in Fig. 1,  $V(q)=\cos(2 \pi q)$ 
and Fig. 2 compares in the mixed phase 
space regime the exact quantum calculation with a purely classical simulation
corresponding to $\sigma^2(k,N)_{cl}$. We see that a simple classical
simulation reproduces the curve extremely well, including the secondary hump,
till around $k \approx 2 \pi$. It is quite remarkable that the classical curve
continues to pick out the initial bessel function oscillations in the 
deeply chaotic regime. In the figure the time average is done over an ensemble
for the first hundred iterations, and the oscillations indicate short time
correlations that will strictly disappear with increasing time.

The transition to classical chaos is accompanied then by a transition of the
variance of the level velocities from a quadratic to a linear $N$ dependence.
Based on this observation we may write a general expression for the
variance of the level velocities as a Weyl series with principal terms
\beq
\label{genexp}
 \sigma^2(k,N)\, =\, c_1(k) \, N\, +\, c_2(k) \, N^2,
\eeq
where $c_1(k)$ and $c_2(k)$ are system dependent and we have given above their
expressions assuming only one of them is appreciable. Note that 
we have not evaluated $c_1(k)$ in the mixed phase-space regime, and that this
will not in general vanish. On the other hand we expect that $c_2(k)$ vanishes
as a classical transition to chaos occurs. This is illustrated in Fig. 3,
where $c_2(k)$ is evaluated based on a best fitting curve using five $N$ 
values, equally spaced, between 100 and 500. The curve is fitted by assuming a 
third-order polynomial in $N$ for which the co-efficient of $N^3$ returned by 
the fit was always of the order of $10^{-6}$ or less.

In general the RMT result derived earlier Eq.~(\ref{rmt3}) will be correct 
under the  assumption that 
$C(l)\, = C(0) \delta_{0,l}$,
implying  delta correlated processes. Thus the departures from universality 
is related to higher order time correlations. The response of the system is 
not only dependent on the strength of the perturbation, but also on the
dynamical correlations inherent to the system.

\subsection{Generalizations}

	We remark now on the general case $\overline{V} \ne 0$. This implies
an overall drift to the energy levels due to changing phase space volumes.
Using Eq.~(\ref{vjvjp}) and adding and subtracting $n \overline{V}^2$ from the 
diagonal part of Eq.~(\ref{modtraceV}), we get after using  
$<|\mbox{Tr}\, (U^n)|^2>_n \, =\, N$,
\beq
\label{gen1}
\sigma^2(k,N)\, =\, \frac{N}{4 \pi^2} \, g \left( D(k)\, -\, \overline{V}^2 \right).
\eeq
The large chaos limit of this is
\beq
\label{gen1rmt}
\sigma^2(\infty,N)\, 
=\, \frac{N}{4 \pi^2} \, g \left( \overline{V^2}\, -\, \overline{V}^2 \right),
\eeq
and is the RMT result.

Variations of two independent parameters is an important problem, considering
that many novel effects, including geometric phases may be observed. Here
we will consider, in a generalization of the above, correlations between
independent parameter variations. We assume that the Level velocities
are given by the matrix elements:
\beq
\label{matele}
\frac{\partial \phi_j}{\partial \lambda_i} 
\, =\, \frac{ N}{2 \pi} 
\br \psi_j |V_i| \psi_j \kt,
\eeq 
$i=1,2$, and $\lambda_i$ are two independent parameters while $V_i$ are two
Hermitian operators. Thus the correlations considered below are also
correlations between diagonal matrix elements of two arbitrary Hermitian 
operators. 

We derive then that
\beq
\label{corrave}
 \frac{1}{N}\sum_{j=1}^N 
\frac{\partial \phi_j}{\partial \lambda_1} 
\frac{\partial \phi_j}{\partial \lambda_2} \, =\,\frac{N}{4 \pi^2} 
\left<  \mbox{Tr}(U^n V_1) \, \mbox{Tr}(U^{-n} V_2) \right>_n.  
\eeq
Using methods as outlined above the correlation function is semiclassically
evaluated and we get
\beq
\label{corroper}
\sigma(\lambda_1,\lambda_2)\, \equiv\,  
\frac{1}{(\sigma_1\, \sigma_2)} \left(
\left< \frac{\partial \phi_j}{\partial \lambda_1} 
\frac{\partial \phi_j}{\partial \lambda_2} \right>_j
\, -\, 
\left< \frac{\partial \phi_j}{\partial \lambda_1} \right>_j
\left< \frac{\partial \phi_j}{\partial \lambda_2} \right>_j \right)
\, \sim\, 
\frac{\left( D(\lambda_1,\lambda_2) \, -\, \overline{V_1}\,   
\overline{V_2}  \right)}{
\sqrt{(D_1-\overline{V_1}^2)( D_2-\overline{V_2}^2)}}
\eeq
The function $D$ is a generalization of Eq.~(\ref{correlD}) 
involving the dynamical
correlation between the functions $V_1$ and $V_2$:
\beq
D(\lambda_1,\lambda_2)\, =\, \frac{1}{{\mathcal A}} \left( \int_{\mathcal A}
V_1(q_0,p_0) V_2(q_0,p_0) \, dq_0 dp_0 \, +\, 2 \sum_{l=1}^{\infty}
\int_{\mathcal A} V_1(q_0,p_0) V_2(q_l,p_l) \, dq_0 dp_0 \right),
\eeq  
the dynamical variables after a time $l$ integrated from $(q_0,p_0)$ is
denoted $(q_l,p_l)$.
While $D_{1,2}$ refer
to the correlations appropriate to them individually and defined earlier
in Eq.~(\ref{correlD}).  
We remark that our derivations have assumed time-reversal symmetry,
and that the factor $g$ is responsible also for phase-space symmetries. 
It may be generalized to include the factors that come due to 
breakdown of time-reversal, or inclusion of spin. 

We finally consider the situation where not only the averages 
$\overline{V}_i$ vanish, but also there is no tangible correlation 
between them, {\it i.e.,} $D(\lambda_1,\lambda_2) \, =\, 0$.
Then the semiclassical expressions above give zero and are
incapable of capturing the small albeit finite and rapid oscillations
(with parameter). This limitation is of course evident all along, including
Fig. 1, where the Bessel function captures the low frequency oscillations
only. 

In Fig. 4, is one such example, where we have considered $V_1\, =\, 
\cos(2 \pi q)$
and $V_{2}\, =\, \cos(4 \pi q)$. The parameter $\lambda_2$ is set as zero so that
the relevant classical system is still the standard map, while the other
parameter is $k$ above. The correlation is seen as a function of this last
parameter, the average actually vanishing. We see that this measure too
captures the transition from mixed phase space to chaotic phase space, but 
that after the transition there are only extremely rapid oscillations 
about zero, although there are quite frequently fairly large correlations.
In fact the frequency of the oscillations are so rapid that 
they seem to have self-similar properties as a random fractal.  
We may estimate the order of magnitude of the frequency if we assume that
these arise from the off-diagonal part of the semiclassical sums. The 
magnitude of the parameter change needed to change a typical orbit action 
by $h$ or $(1/N)$ is needed. From the fact that action changes have a variance
proportional to the period, we get $|\Delta S| \sim \sqrt{n} |\Delta \lambda|$,
and therefore $|\Delta \lambda| \sim N^{-3/2}$, if we take as the 
period $n =N$, which is the Heisenberg time and represents the time by which 
the spectrum is practically resolved.

In conclusion then, we have studied variances of level velocities and their
generalizations in the chaotic as well as mixed phase space regimes.
Noting that the transition to chaos
is perfectly reflected in this measure, we derived detailed formulae for
them, in terms of classical correlation coefficients and illustrated this
with the help of the standard map. 
The mixed phase space regime was surprisingly well captured by a simple 
classical estimate. The observations of oscillations or variations
in the level velocity variances due to classical correlations,
as well as using them to distinguish mixed from chaotic phase space are
both experimentally accessible.

The possibility of using the level 
velocity in conjunction with wavefunction intensities in a measure of 
phase-space localization has been proposed \cite{Steve}, 
and it is hoped that this detailed understanding of the level velocities
will help in this as well. In particular this measure is a 
special case of the correlation between two operators discussed above, with
an important complication being that the Wigner transform of the
relevant operator, a projector in phase space, varies over scales of
order $\hbar$. We have also noted that the RMT results, after 
assuming independence of eigenvalues and eigenfunctions is 
capable of predicting the level velocities only in the limit of 
extremely large chaos, or equivalently ignoring all higher order time
correlations other than the zeroth.

This work was supported by NSF-PHY-9800106 and the ONR.

\begin{center}
{\bf Figure Captions}
\end{center}

{\bf Figure 1} Scaled variance as a function of the parameter $k$, 
N=300, a=0.35. The inset shows a part of the plot magnified, the points
are numerical data while the smooth line is the twice the diffusion 
coefficient.

{\bf Figure 2} Same as Fig. 1, comparing, mainly 
in the mixed phase space regime,
the exact variance (dots) with the classical estimate (the line). 
The classical estimate is after averaging over a hundred time steps.          
  
{\bf Figure 3} The coefficient $c_2(k)$ as a function of the parameter; 
note that it reflects the classical transition to chaos.

{\bf Figure 4} The quantum  correlation between the  matrix elements of the two operators
$\cos(2 \pi q)$ and $\cos(4 \pi  q)$
whose classical correlation vanishes in the chaotic regime.  
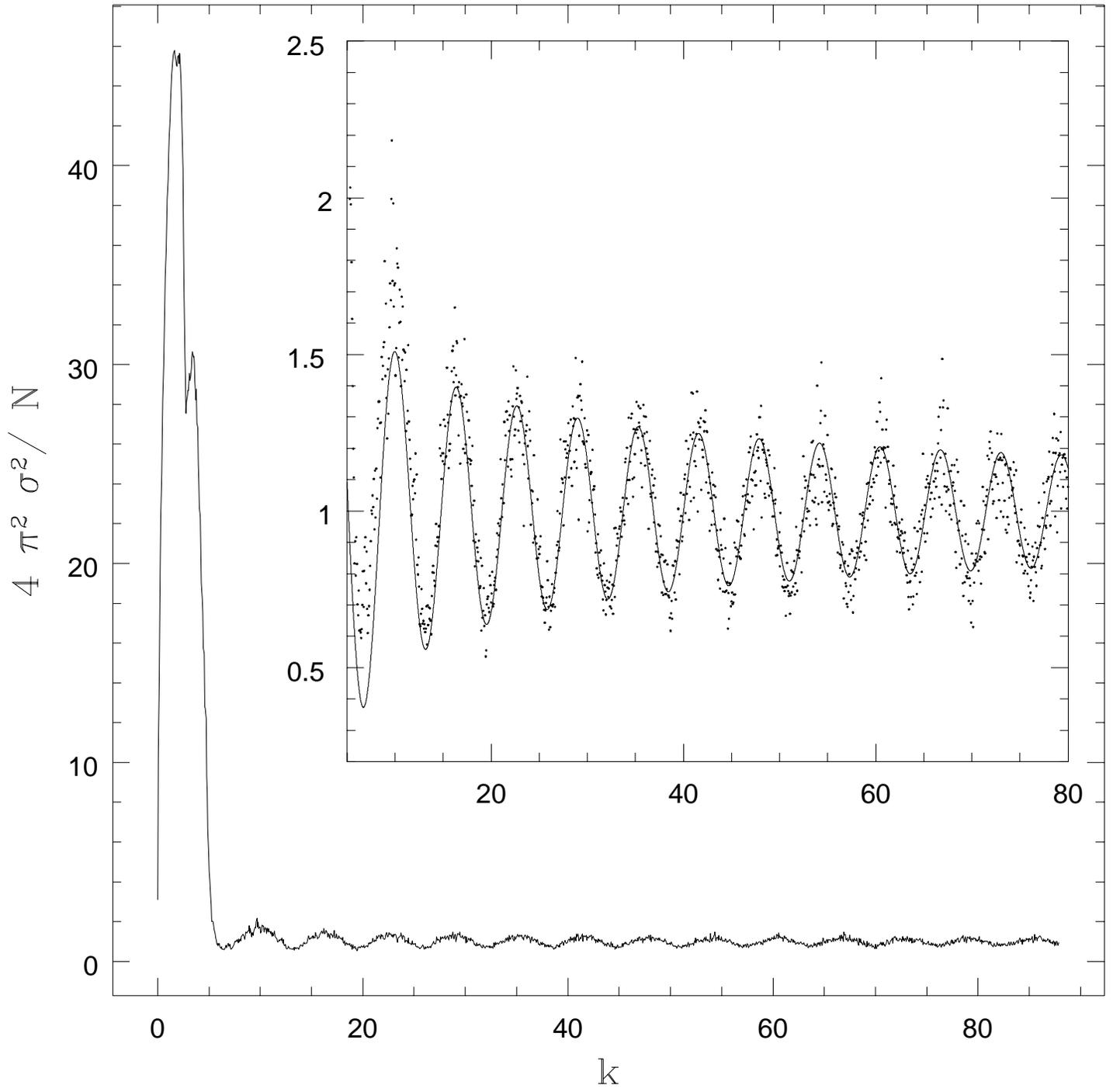
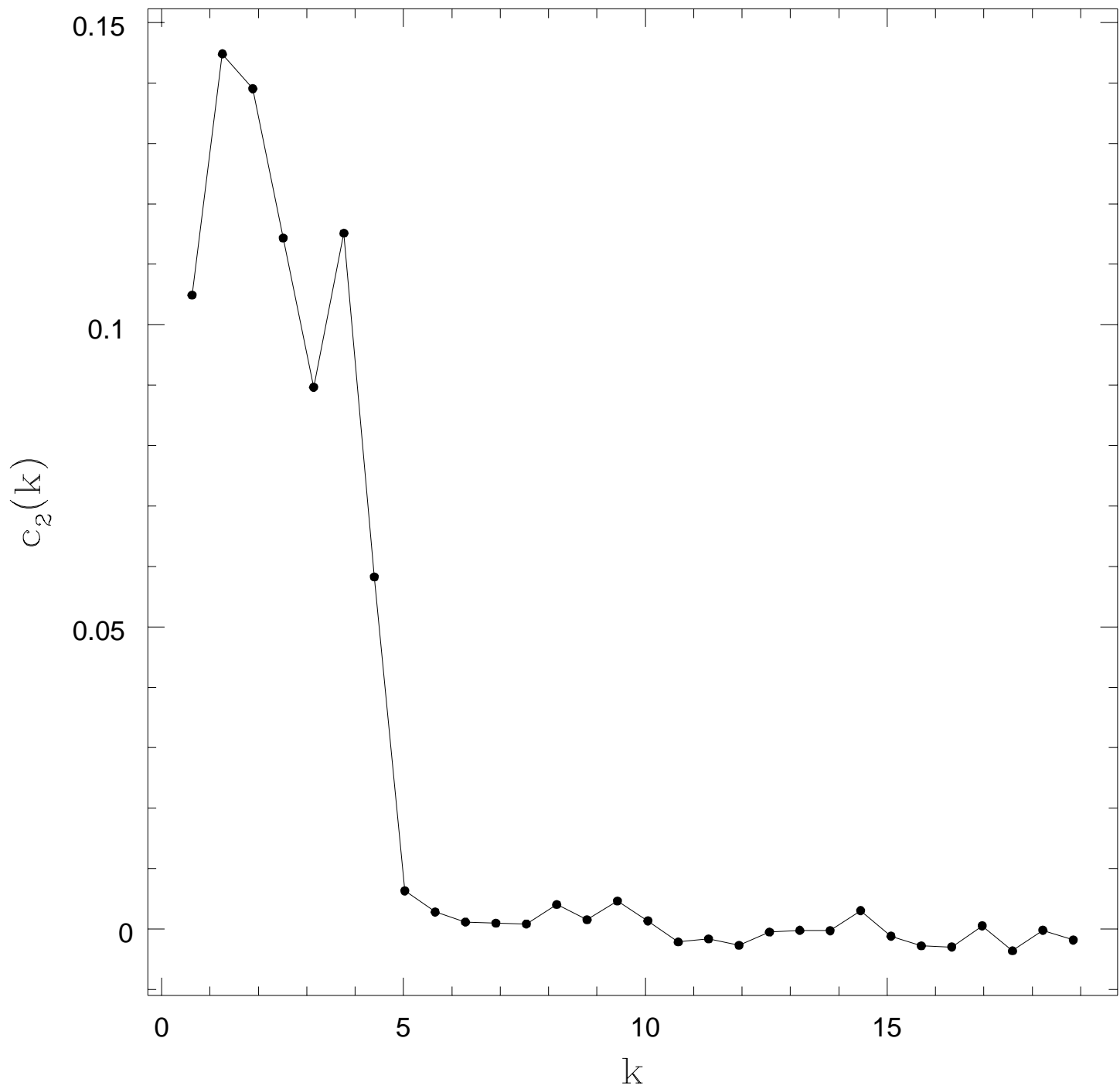
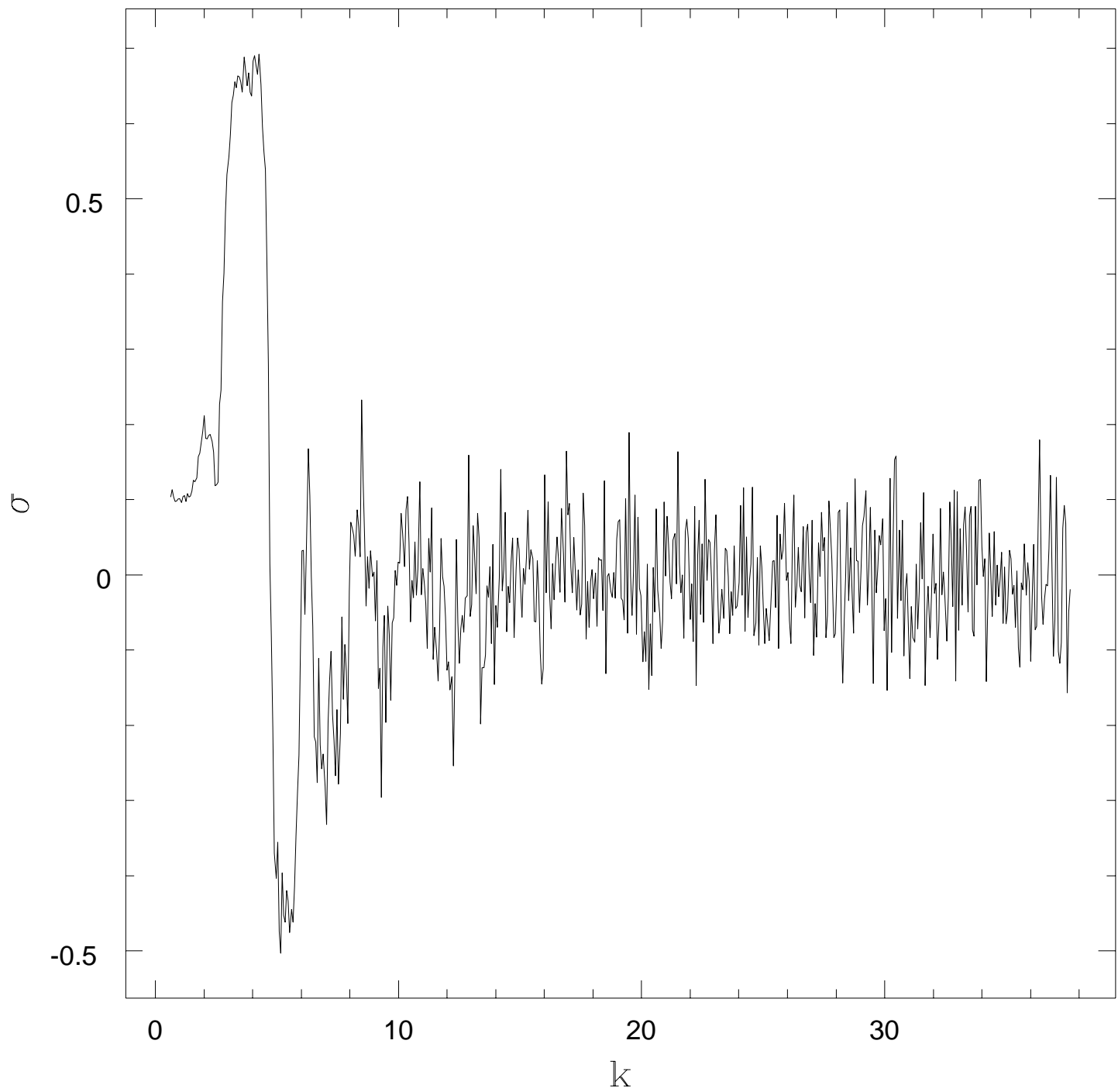
\end{document}